\documentstyle[12pt]{article}
\newcommand{\ssy}[5]{#1 19#4 {\it #2\/} {\bf #3} #5}
\newcommand{\rnum}[1]{$\rm I\!R^#1$}
\newcommand{\pn}{\par\noindent}
\begin{document}
\title{Time machines with non-compactly generated Cauchy horizons and
``handy singularities".}
\author{S. V. Krasnikov\thanks{E-mail: redish@pulkovo.spb.su}}
\maketitle
It is 10 years now that time machines (TM's) are intensively studied,
but the main question, of {\em whether or not TM's can be created} remains
still unanswered. 
\par
Most investigations so far have centered on
TM's with compactly generated Cauchy horizons (compact TM's, or
CTM's, for brevity) \cite{Conj}. It has been understood that
creation of such TM's is connected with at least two serious problems:
\pn{\it (1)\/} There are some ``dangerous" null geodesics in the causal
regions of CTM's. A photon propagating along such a geodesic would
return infinitely many times (each time blue-shifted) in the vicinity
of the Cauchy horizon \cite{Clas}. This suggests that quantum effects could
prevent the creation of the TM. It is still not clear whether they
will \cite{Yur,Quan}, but they might.
\pn{\it (2)\/}  Creation of a CTM inevitably involves violations of 
the Weak Energy Condition. In this connection it is common to refer
to quantum effects, but here again some restrictions exist \cite{QI}.
\par
There is a little hope that any of these issues will be completely
clarified in the foreseen future, since they both involve
QFT in curved background which has a lot of its own unsolved problems.
\par 
It well may be, however, that actually we need not clarify them.
Indeed, {\em why must we bound ourselves to compact TM's}?
The only answer I  met in the literature is that
noncompactness implies that either 
infinity or a singularity \cite{Ori} are involved. So, ``extra
unpredictable information can enter the spacetime" \cite{Conj}
and we can no more completely control such a TM. 
This is true, indeed, but the point is that this in no way is the
distinctive feature of ``noncompact" TM's (NTM's). In {\bf any} time
machine we 
encounter extra unpredictable information as soon as we intersect
the Cauchy horizon (by its very definition) and so for {\bf none} of TM's the
evolution can be completely controlled from the initial surface.
\par 
So, we conclude that NTM's are not a bit worse than CTM's and
preference given to the latter is a sheer matter of tradition \cite{Ori}.
Meanwhile, the example of the Deutsch-Politzer spacetime \cite{Deu} (DPS)
shows that to create an NTM we need neither ``dangerous geodesics",
nor ``exotic matter". (Of course the problem remains of how {\em to cause}
a spacetime to evolve in the appropriate manner but as noted above
this does not depend on compactness.) 
\par
The DPS is obtained as
follows. A cut is made along a spacelike segment (i.~e.\ disk $D^1$)
on the Minkowski plane. A copy of the cut is made to the past from the
original one. The boundaries of the segments (i.~e.\ 4 points, or two
copies of $S^0$) are removed from the spacetime and the banks of the
cuts are glued, the upper bank of each cut is glued to the lower bank
of the other cut. The four removed points cannot be returned back and
form thus irremovable singularities.
\par
What enables us to render the abovementioned
theorems harmless in the case of the DPS is the
presence of these quite specific singularities and what makes them so handy
is the following 
	\begin{enumerate} 
	\item Unlike  Misner-type singularities, 
	they make the relevant region noncompact, 
	\item They are absolutely mild (i.~e.\ all 
	curvature scalars are bounded) and so there is no need to invoke
	quantum gravity to explore these singularities,
	\item And they are of laboratory, rather than of cosmological nature.
	That is they are confined in such a region $R$ of the spacetime $M$
	that $M-R=\;$(a ``good" spacetime)${}-{}$(a compact set).
	\end{enumerate} 
All the above suggests that
singularities possessing these properties (we shall call them
{\em handy singularities}) are worth studying. 
\par
Twenty years ago Ellis and Schmidt \cite{Mild} constructed several
singularities satisfying (1) and (2), but not (3). Besides, those
singularities 
occuring in Minkowski spaces quotiented by discrete isometries
were too symmetric, which was interpreted as instability.
\par Recentely an $n$-dimensional analog of the DPS
 was  obtained \cite{Gen} by the replacement:
	\begin{equation}
	D^1\;\to\; D^{n-1}, \qquad S^0\;\to\; S^{n-2}
	\label{repl}
	\end{equation}
in the procedure described above.
So, we know that $n$-dimensional handy singularities exist, but that
is all we know. 
\par
And now I would like to propose a simple trick (just generalizing
that from \cite{Gen})  for constructing 
quasiregular singularities (including the previously known), which
yields at the same time a variety of handy singularities.
			
\paragraph{1.} Take a region $R$ in an $n$-dimensional spacetime $M$ and an
$(n-1)$-dimensional submanifold $S\subset R$ bounded by a closed
$(n-2)$-dimensional submanifold $C$. Denote by $\widetilde M$ the
universal covering of $M-C$  and by $\pi$ the natural projection 
$\widetilde M \mapsto M-C$. Let $p$ be the ``natural embedding"
$p: M-C\mapsto\widetilde M ,\; p=\pi^{-1}$.

\paragraph{2.} Now make a cut along $S$ in $M$, that is consider 
 $M_S\equiv\overline{p(M)}$. 
Note that $S_0\equiv M_S-p(M)$ is a double covering of
$S$: $\pi(S_0)=S$.  If $S$ is orientable, $S_0$ is just a disjoint
union of two copies of $S$ (the two ``banks" of the cut): $S_+$ and
$S_-$. The projection $\pi$ induces a nontrivial isometry $\sigma: S_0\mapsto
S_0$ enabling one to return to $M$ from $M_S$ by ``gluing the banks".
 Namely, $M_S/\sigma=M-C$

\paragraph{3.} As the third step take an  isometry $\eta:\:R\mapsto R'$,
and repeat the above procedure with $M$ replaced by $M_S$ and $S$
replaced by $S'\equiv\eta(S)$. The resulting space $M_{SS'}$ is $M$
with two cuts 
 made (along $S$ and along $S'$), each taken with its ``banks".
The desired spacetime $N$ can be obtained now by the appropriate
identification:
	\begin{equation}
	N \equiv M_{SS'}/ \xi
	\label{ide}
	\end{equation}
where $\xi\equiv\sigma\circ\eta$ (rigorously speaking instead of
$\eta$ we should have written some $\eta'$ in (\ref{ide}), where
$\eta'$ is the continuous extension of $p\circ\eta$ on $M_{SS'}$,
we shall neglect such subtleties for simplicity of notation). In the
orientable case (\ref{ide}) simply means that we must glue $S_\pm$ to
$S_\mp$. 
\par
When $\eta$ is nontrivial, we cannot 
return $C$ back and $N$ contains thus a handy singularity ``in the
form of" $C$. We can also use in (\ref{ide}) any other isometry
$\xi\neq\sigma$ (see example (d) below).
\par 

If $S$ was chosen to be a disc $D^{n-1}$ in the Minkowski space and
$\eta$ to be a translation, we obtain the
DPS (cf.\ (\ref{repl})), but being applied to different $M,\,S,$ and
$\eta$ the same procedure will give us a lot of quite different
``handy singularities".

\paragraph{3-dimensional examples.} In what follows $M$ for simplicity
is taken to be a flat space \rnum{3} and
$z,\;\rho,\;\phi$ are the cylinder coordinates in it.

\pn{\bf (a)} Let
$C$ be an arbitrary knot (with $S$ being its Seifert surface) and $\eta$ be a
translation. Then $N$ is what is called a ``loop-based wormhole" in 
\cite{Vis}. 

\pn{\bf (b)} Let $S$ be a disk $z=0,\;\rho < \rho_0$ and $\eta$ be a
rotation $\phi \to \phi+\phi_0$ (note that $S=S'$ thus). In this case $N$
has quite a curious structure.  It is diffeomorphic to $M-C$, any
simply connected region of $N$ is isometric to some region of $M-C$
and vice versa, and still globally they are not isometric.
Similarly to the conical case one can think of $N$ as a
space \rnum{3} having delta-like Riemannian tensor with support in $C$ (in
contrast to $M$, where the Riemannian tensor is zero even in the
distributional sense). 

\pn{\bf (c)} Take a rectangle strip. Turn one of its ends through
$n\pi$ and then glue it to the other (so that a cylinder is obtained
if $n=0$ and the M\"obius band, if  $n=1$). Take the resulting surface for
$S$.  $S$ can be specified by condition
	$$ 0 < r < 1/2, \quad \theta=\pm n\phi $$
Here $r$ and $\theta$ are new coordinates. For any point $A$, $ r(A)$
is defined to be
the distance from $A$ to $a$ and $\theta(A)$ to be the angle between
the $z$-axis and the direction (Aa), where $a$ is the point
$z(a)=0, \; \rho(a)=1, \; \phi(a)= \phi(A) $. 
For a (local) isometry one can take $\eta:\;\phi \to \phi+\phi_0,\; 
\theta  \to \theta + n\phi_0$.
When $n=3$, $C$ is a knot (trefoil).

\pn{\bf (d)} Now change \rnum{3} to \rnum{3}${}\setminus\{0\}$ in example (b) 
and take $\xi$ in (\ref{ide}) to be the reflection
${\bf r} \to -{\bf r}$. Thus
obtained $N$ has a 
singularity generated by the circle $C$ and two more corresponding to
the removed $\{0\}$ (it also cannot be returned back). The latter
though being handy singularities have
locally the same structure as a singularity considered in \cite{Mild}.

\end{document}